# Analogue Sum ASIC for L1 Trigger Decision in Cherenkov Telescope Cameras

**Joan Abel Barrio**[a], **Oscar Blanch**[b], **Joan Boix**[b], **Eric Delagnes**[c], **Carlos Delgado**[d], **Lluís Freixas Coromina**[d*], **David Gascón**[e], **Fabrice Guilloux**[c], **Ruben López Coto**[b], **Gustavo Martínez**[d], **Andreu Sanuy**[e] and **Luis Ángel Tejedor**[a]

[a] *Universidad Complutense de Madrid (UCM),*
*Ciudad Universitaria, Plaza Ciencias, s/n, 28040, Madrid, Spain*

[b] *Institut de Física d'Altes Energies (IFAE)*
*Edifici CN, Campus UAB, 08193, Bellaterra, Spain*

[c] *Commissariat à l'énergie atomique (CEA-SACLAY)*
*91191 Gif-Sur-Yvette, Cedex, France*

[d] *Centro de Investigaciones Energéticas, Medioambientales y Tecnológicas (CIEMAT)*
*Av. Complutense, 40, 28040, Madrid, Spain*

[e] *Universidad de Barcelona (UB),*
*Martí i Franquès, 1, 08028, Barcelona, Spain*

*E-mail*: lluis.freixas@ciemat.es

ABSTRACT: The Cherenkov Telescope Array (CTA) project [1] aims to build the largest ground-based gamma-ray observatory based on an array of Imaging Atmospheric Cherenkov Telescopes (IACTs). The CTA will implement a multi-level trigger system to distinguish between gamma ray-like induced showers and background images induced by night sky background (NSB) light [2]. The trigger system is based on coincident detections among pixels (level 0 trigger), clusters of pixels (level 1) or telescopes. In this article, the first version of the application specific integrated circuit (ASIC) for Level 1 trigger system is presented, capable of working with different Level 0 strategies and different trigger region sizes. In addition, it complies with all the requirements specified by the CTA project, specially the most critical ones as regards noise, bandwidth, dynamic range and power consumption. All these features make the presented system very suitable for use in the CTA cameras and improve the features of discrete components prototypes of the L1 trigger circuit in terms of compactness, noise, performance and power consumption.

---

[*] Corresponding author.

# Contents

## 1. Introduction

Cherenkov Telescopes are used in gamma-ray astronomy for the detection of cosmic rays after their interaction with earth's atmosphere [1] [3] [4] [5]. By detecting and processing Cherenkov light produced by gamma rays of cosmic origin, it is possible to characterize the incoming photon in terms of direction and energy. Placing several telescopes (of different sizes) together results in a gamma-ray detector with enhanced sensitivity in a wide energy spectrum.

Detection of Cherenkov light is performed by means of cameras whose pixels are based on fast and sensitive photo sensors; typically photomultiplier tubes (PMTs). Each photo sensor is coupled to an electronic readout chain, in order to digitize the PMT waveform. An electronic trigger chain for event detection is used to optimize resources in terms of memory and data bandwidth.

Trigger decision electronics perform a fast signal processing in order to discriminate gamma-ray events from other events produced by light from Night Sky Background (NSB) [2]. This contribution presents a new analogue sum Application Specific Integrated Circuit (ASIC) designed specifically for camera trigger decision in Cherenkov Telescopes. Section 2 describes a multilevel approach used for this type of trigger systems. Section 3 presents the architecture of a new ASIC for the implementation of the first level of the trigger. Section 4 includes measurements of functionality and performance of the ASIC. Conclusions are included in section 5.

## 2. System Architecture of the Trigger

### 2.1 Analogue Trigger Concept

The trigger decision in Cherenkov telescopes is based in the detection of a concentration of signal both in space and time. The photons coming from a Cherenkov shower will be detected by several close pixels at nearly the same time. Therefore, a telescope trigger, requiring a given number of pixels of the camera with a minimum number of photons in a narrow time window, will reject a sizeable amount of NSB induced events.

The analogue trigger concept takes full advantages of the analogue signal from pixels in the cluster in order to select the meaningful information of the smaller contributions for the purpose of decreasing the threshold and improving the sensitivity. A two-level trigger scheme compatible with the actual hardware and mechanical architecture of telescope cameras for CTA has been proposed.

Mainly, there are two different analogue trigger strategies. The so-called Majority trigger architecture consists of the discrimination of signals coming from pixels, in order to count the number of activated pixels in a region. Alternatively, the so-called Sum trigger strategy uses the addition of the analogue pulses from the pixels in a region, in order to compare the result of their analogue sum with a threshold defined for the whole region. In a typical multilevel architecture, these two strategies are used at the level 0 (L0) trigger, being possible to use the same level (L1) trigger circuit for both. This trigger decision scheme is well documented in previous papers [6].

### 2.2 Architecture of the Trigger

The trigger system is designed for cameras made of 7-pixel clusters. Each cluster consists of PMTs, front-end circuits, digitization and readout electronics and trigger decision and distribution electronics [6]. The signal of every pixel is amplified in the front-end board and feeds a fast digitizer mainly based on NECTAR [7] or DRAGON [8] developments.

The first trigger level [9], so-called L0, is cluster-based and combines the signals of the pixels in each cluster. In this level, the trigger implements the Majority and Sum trigger strategies. The second level, called L1 [10], is also implemented in each cluster and combines the L0 signals of neighbouring clusters with local L0 in specific trigger regions. The level 1 will generate a digital trigger pulse, which is distributed to the digitizers of all clusters in the camera and is the final camera trigger. A detailed description of the ASIC implementing L1 subsystem is presented in section 3.

## 3. ASIC L1

A mixed signal ASIC has been specifically developed, implementing the functionality of analog level 1 trigger decision prototype [10] for CTA Cherenkov Telescope cameras [1]. The ASIC comprises 7 input differential analogue channels and 2 output digital differential channels. Analogue inputs are provided by the previous trigger stage implemented in the so-called L0 ASIC [9]. The two basic functionalities of the L1 ASIC are the calculation of the sum of three configurable sets of inputs channels and the discrimination of the resulting voltage pulses in order to generate digital trigger output signals when any of the sums is above configurable voltage thresholds. The analogue signal processing stage has been specifically developed for this application by means of a low noise differential architecture that provides an estimated bandwidth of 500 MHz.

### 3.1 Specifications

On one hand, there are a set of previous specifications defined mainly by the PMT signals that have to be processed in order to fulfil the physics requirements of CTA, in particular, signal-to-noise ratio (SNR) and bandwidth. On the other hand, the goal is to replicate the functionality and improve the performance of the L1 trigger system implemented with discrete components on a mezzanine board [10].

### 3.1.1 Flexibility

Apart from the fact that, L1 ASIC is able to work with different trigger strategies (Majority and Sum trigger), it is able to operate in the three working modes defined in L1 discrete components version. Corresponding to trigger regions of 2, 3 or 4 clusters (14, 21 or 28 pixels), allowing the trigger to achieve optimized sensitivity in different sky brightness conditions. The patterns of region additions in the three different modes achieve an overlap across the camera in such a way that trigger regions are uniform in the whole camera without redundancy. This concept is well explained in Figure 3 of reference [10]. Moreover, the L1 ASIC provides more flexibility and uniformity, since any input combination can be processed and all input channels have the same architecture, as is shown in Figure 1.

### 3.1.2 Noise

The noise after the addition of signals is specified to be lower than 0.2 photoelectrons (phe) per added channel. This is especially critical for the L0 Sum trigger strategy, since there is no discrimination at that stage, and therefore small contributions of signals with low signal-to-noise ratio are summed up in the L1 ASIC.

### 3.1.3 Dynamic Range, Linearity and DAC Resolution.

The trigger system must be able to process at least between 0.2 and 100 phe. As each photoelectron corresponds to 10 mV at the L1 input, the signal range is specified from 2 mV to 1000 mV. The gain of the different stages inside the ASIC has been chosen to fulfil a trade-off among linearity, signal-to-noise ratio and threshold range and resolution. This gain is defined to avoid saturation when the entire signal is concentrated in one channel and to keep linearity error low in the full range. In addition, the Digital-to-Analogue converters (DAC) range is matched to the input signal range in order to optimize threshold resolution: two 9-bits DACs provide a resolution of about 0.2 phe. Finally, internal gain provides a minimum signal-to-noise ratio of 5.

### 3.1.4 Bandwidth

The bandwidth of the L1 ASIC is specified to be greater than 500 MHz in order to ensure that a pulse of 3.25ns is correctly processed. The input stage fulfil widely the bandwidth specification, being the adder the limiting stage of analogue processing part in terms of bandwidth.

### 3.1.5 Power Consumption and Compactness

In CTA camera designs, all electronics are installed at the focal plane [1]. This fact imposes hard restrictions in terms of compactness, power consumption and weight. Hence, any design which reduces one or more of these parameters is worth pursuing. The power consumption of the L1 ASIC is less than 350 mW, less than 50% of the power consumption of the discrete components version (800mW). A compact design is very suitable in order to save space and provide improved integration in the readout system. This has also advantages in terms of reliability. It is also worth pointing out that the cost of an ASIC development is justified in CTA due the high number of channels in the experiment.

### 3.2 Architecture of the L1 ASIC

Figure 1 shows a basic block diagram of the ASIC. The first version of the analog trigger L1 ASIC comprises seven input differential analogue channels and two output differential digital channels. The two basic functionalities of the circuit are the calculation of the analogue sum of the input channels and the discrimination of the resulting voltage pulse in order to generate digital trigger L1 output signals. Input pulses are replicated and connected to three analogue adders through a set of high bandwidth switches, thus, any group of signals can be summed up in any of the three adders. The output of the adders are connected to two groups of discriminators, with independent threshold for each group, in order to generate high level and low level threshold triggers. The threshold voltages are generated by two independent differential DACs. The outputs of the discriminators are combined in two OR gates. The OR's outputs are connected to Low-Voltage Differential Signalling (LVDS) transmitters that provide the digital trigger outputs of the ASIC.

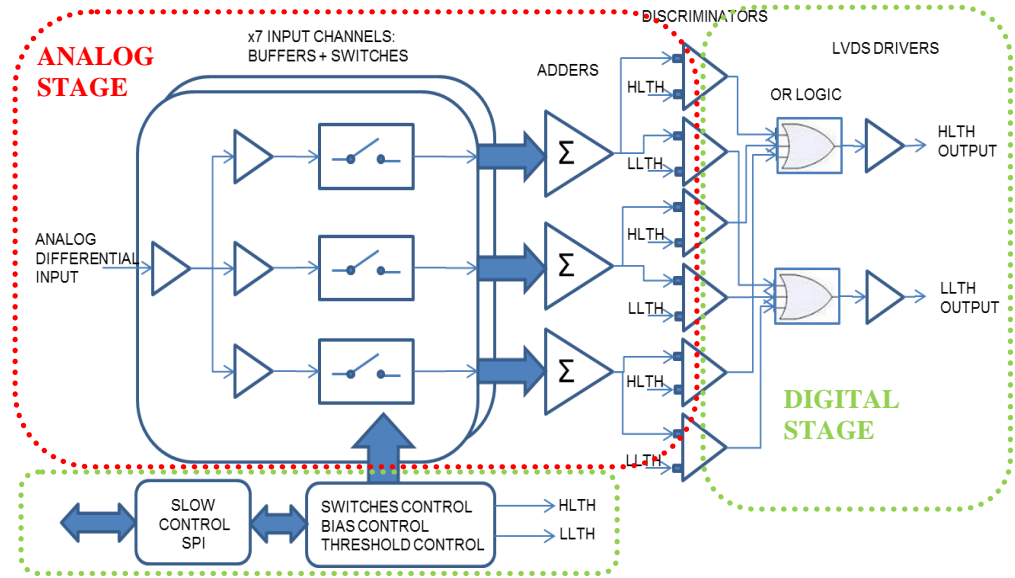

*Figure 1: Block diagram of L1 ASIC*

### 3.2.1 Analogue Stage

A fully differential, low noise and high bandwidth architecture has been designed for all circuits in the analogue signal processing stage of the ASIC, from inputs pins to discriminators inputs, which has been specifically developed for this application. All the blocks of the analogue stage presented below must achieve the specifications detailed in section 3.1.

#### 3.2.1.1 Input Stage and Switch

All channels are replicated three times and connected to one input of each analog adder through a low impedance differential high bandwidth switch. Thus, there are 21 switches, which are controllable in an independent way by the slow control circuit. This symmetric architecture provides high flexibility for defining different trigger regions and ensures all input channels are processed in the same way inside the ASICs in order to minimize differences from channel to channel. The overall gain of the input stages is near 1 and the linear range is around 1 V. The bandwidth of the input buffers and switches has been designed to be much greater than 500MHz.

#### 3.2.1.2 Adder

The ASIC comprises three independent fully differential 7-input analog adders. The overall gain of the adder is ~ 1 and the linear range is around 1 V. The bandwidth of the adder has been designed carefully to be greater than 500MHz because is a very critical block for the bandwidth of all the analog circuit.

### 3.2.2 Digital Part

#### 3.2.2.1 Discriminator

The output of the three adders are connected to two group of discriminators, with independent threshold definition each group, in order to generate a High Level Threshold Trigger and a Low Level Threshold Trigger. Six differential leading edge voltage discriminators are used for trigger signal generation, one per adder and per voltage threshold. The discriminator circuit has been provided by CEA-Saclay [11]. The transition time of the digital pulse is around 1ns and the minimum repetition time, or double trigger resolution, is 5ns.

#### 3.2.2.3 Digital-to-Analog Converter (DAC)

The threshold voltages are generated by two independent differential voltage DACs, with 10 bits resolution, which are controllable by slow control. The individual DAC circuit has been also provided by CEA-Saclay [11]. The reference voltage circuit of the DACs has been modified to match the operational range of the adder in order to optimize the resolution in the threshold definition, obtaining around 0,2 phe resolution in the 100 phe range, and therefore similar to the noise level of the analogue circuit.

#### 3.2.2.3 Output Stage

The outputs of the discriminators are combined in two OR gates and then sent to LVDS transmitters, which provide the 2 digital trigger outputs of the ASIC with independent threshold definition. The behaviour of this stage does not limit the performance of the discriminator outputs.

#### 3.2.2.4 Slow Control

The slow control circuit is based in a serial link. The L1 ASIC acts as slave and needs a Master to operate, typically an FPGA. The Slow control comprises a set of seven 16-bit wide registers that can be written and read by the link. The registers contain the data used to control the input stage switches enabling, the DAC used to define the bias current of the differential pairs in the ASIC, the DACs for voltage threshold generation and the enable of LVDS transmitters.

### 3.2.3 Full Layout

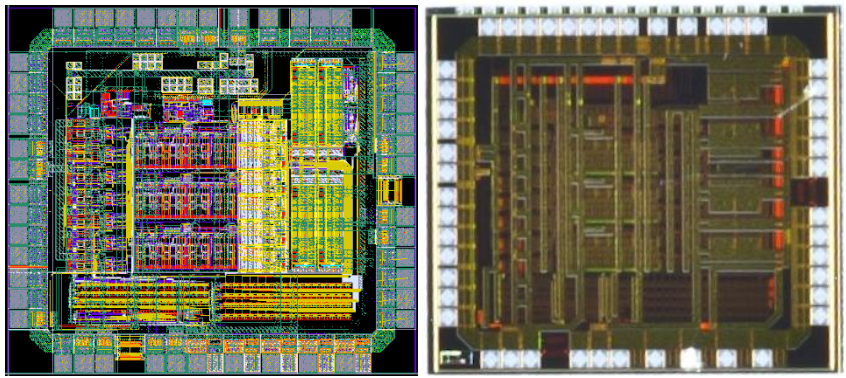

*Figure 2: a) Layout of the L1 ASIC b) die photography of the ASIC*

A first version of the ASIC was submitted in June 2013 for manufacture in a 350nm Silicon-Germanium (SiGe) Bipolar CMOS (BiCMOS) Multi Project Wafer (MPW) run 3.3 LVCMOS by Austria Micro Systems (AMS) to be consistent with other ASICs also developed for trigger and signal adaptation for CTA cameras, for example the L0 trigger of CTA [9], ACTA [12] and PACTA [13]. The dies are packaged in a QFN48 package. 30 units of the ASIC were received in September 2013. The size of the die layout is 2.061 x 1.861 mm, resulting in an area of 3.853 mm$^2$. It has different input/output test pins to test the ASIC properly. Functionality results and preliminary performance results are shown in next section. Figure 2a shows the layout and figure 2b shows the die photography of the ASIC. The ASIC will be soldered in a mezzanine board to be plugged in the front-end board (FEB) prototypes. In the final version of the cluster the L0 and L1 ASIC will be integrated in the FEB.

Figure 3 shows a picture and a block diagram of the test set-up. It includes a mezzanine board with a low inductance 48 pin QFN socket, which permits an easy exchange of the ASIC units to be tested. This mezzanine is plugged into a base board that provides a *fanout* for pulse injection and interfaces with a pulse generator, slow control circuit, power supplies and trigger outputs. An FPGA module is used as a bridge with a computer for slow control. It also implements counters for trigger output characterization. The setup is completed by the signal injection generator, oscilloscope and power supplies.

The initial step of the test includes DC levels, power consumption, slow control and functionality checking in order to verify the basic operations. For ASIC response characterization, a rate scan methodology is used. Rate scanning consists of the comparison between the Signal Generator trigger out rate and the ASIC trigger out rate. This computation is performed with scalers implemented in the FPGA board. This allows us to evaluate the transfer functions, the sum of different channels, the estimated bandwidth and other useful tests.

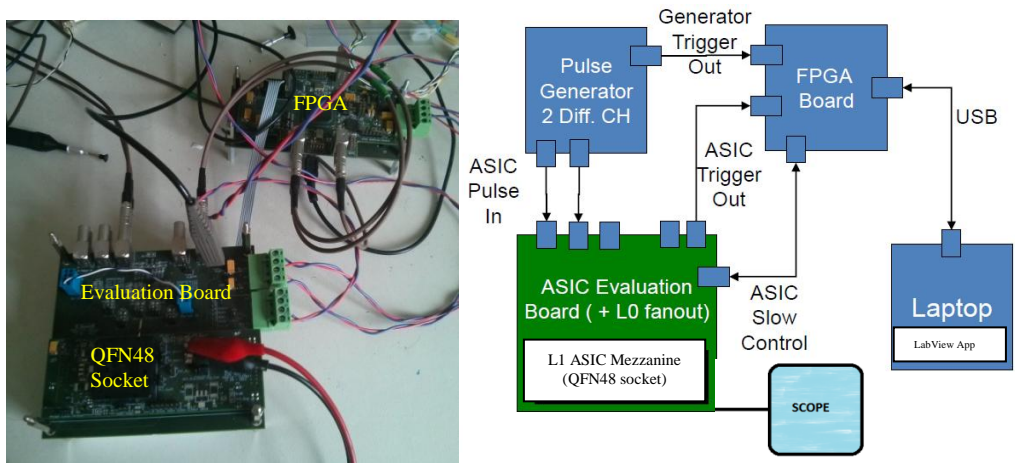

*Figure 3: a) Photography of the set up test boards b) Block diagram of the set up test*

## 4. ASIC Characterization

A full characterization of the 30 ASIC is already finished with very satisfactory preliminary results.

### 4.1 Performance

Figure 4a is obtained with rate scan methodology. It shows a measurement of the mean transfer function of 10 ASICs with linearity error below 10% in a dynamic range of 100 phe. The mean transfer function is obtained with the mean of all different channels and adders combinations transfer functions.

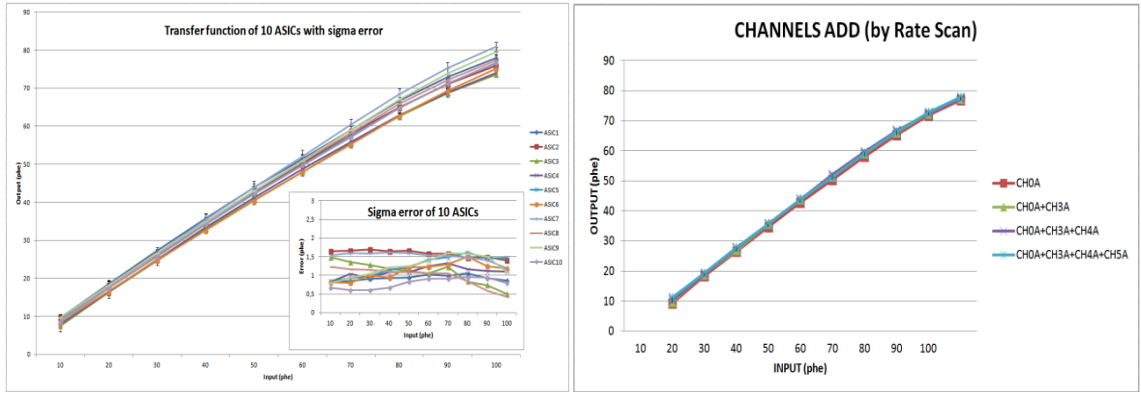

*Figure 4: a) Measurement of the transfer function of 10 ASICs, showing the dynamic range with linearity better than 10%.b) Transfer functions of different Sum combinations by rate scans methods.*

The transfer function of different sum combinations obtained by rate scan is shown in Figure 4b. It can be observed that the gain is compatible in any combination, when the signal is injected in just one channel or distributed in 2, 3 or 4 different channels.

Bandwidth characterization is not trivial since it is not possible to evaluate the amplitude of the signal at the discriminator input in a direct way. Again, a rate scan method is used to measure the gain of the analog circuit. Bandwidth is estimated by changing the input pulse time width and measuring the amplitude of the output signal by rate scan. Figure 5 shows the gain for several pulse time widths. It can be observed that 3 ns pulses are processed without gain loss compared with 10 ns pulses. A -3dB cut-off is obtained for pulses below 1.7ns. This result confirms that the ASIC is able to process the pulses expected in CTA operation without amplitude distortion.

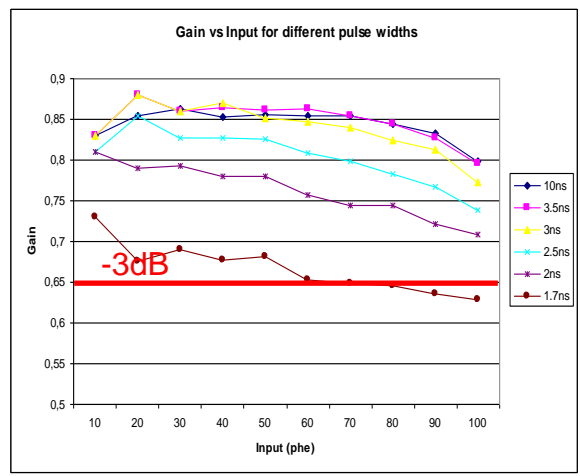

*Figure 5: Gain vs. Input for different pulse widths signals with less than -3dB drops.*

Measured power consumption is around 330mW @ 3.3 V power. Regarding timing, both jitter and propagation delay is measured directly with an oscilloscope. The measured jitter is around 20 ps (RMS), depending on threshold and overdrive. Finally, the delay is 5-6 ns, depending on threshold.

# 5. Conclusions

A first version of an ASIC has been designed and developed for analog sum based trigger decisions for future Cherenkov telescope cameras with the following features: fully differential, low noise, high bandwidth, 1V (100 phe) dynamic range, multi channel, good linearity and flexibility. The ASIC was submitted to a 350nm SiGe BICMOS Multi-Project Wafer run last June 2013 and has been received and tested. The results of the full characterization of 30 ASICs are very satisfactory.

**Acknowledgments** This work has been supported by Ministerio de Economía y Competitividad, project FPA2010-22056-C06-03